\documentclass[conference]{IEEEtran}
\IEEEoverridecommandlockouts
\usepackage{cite}
\usepackage{amsmath,amssymb,amsfonts}
\usepackage{algorithmic}
\usepackage{graphicx}
\usepackage{textcomp}
\usepackage{xcolor}

\usepackage{algorithm}
\usepackage{algorithmic}
\usepackage{xcolor}
\usepackage{tabularx}
\usepackage{subcaption}
\usepackage{cite}
\usepackage{url}
\usepackage{amsmath,amssymb,amsfonts}
\usepackage[numbers]{natbib}
\usepackage{algorithmic}
\usepackage{graphicx}
\usepackage{textcomp}
\usepackage{xcolor}
\usepackage{booktabs}

\usepackage{xspace}
\usepackage{csquotes}
\usepackage{xcolor}
\usepackage{multirow}

\newcommand{\peixian}[1]{\textbf{\textcolor{blue}{peixian: #1}}}

\newcommand{\one}{({\em i}\/)\xspace}
\newcommand{\two}{({\em ii}\/)\xspace}
\newcommand{\three}{({\em iii}\/)\xspace}
\newcommand{\four}{({\em iv}\/)\xspace}

\def\eg{\emph{e.g.}\xspace}

\def\ie{\emph{i.e. }\xspace}

\newcommand{\pb}[1]{\vspace{0.75ex}\noindent{\bf \em #1}\hspace*{.3em}}

\def\BibTeX{{\rm B\kern-.05em{\sc i\kern-.025em b}\kern-.08em
    T\kern-.1667em\lower.7ex\hbox{E}\kern-.125emX}}
\begin{document}

\title{The Emergence of Threads: The Birth of a New Social Network\\}

\author{
\IEEEauthorblockN{
Peixian Zhang, 
Yupeng He,
Ehsan-Ul Haq, 
Jiahui He,
and Gareth Tyson}
\IEEEauthorblockA{Hong Kong University of Science and Technology (Guangzhou)}
Email: \{pzhang041, yhe382\}@connect.hkust-gz.edu.cn\quad \\ euhaq@hkust-gz.edu.cn\quad\ jhe976@connect.hkust-gz.edu.cn\quad\ gtyson@ust.hk
}

\maketitle

\begin{abstract}
Threads, a new microblogging platform from Meta, was launched in July 2023. In contrast to prior new platforms, Threads was borne out of an existing parent platform, Instagram, for which all users must already possess an account.
This offers a unique opportunity to study platform evolution, to understand how one existing platform can support the ``birth'' of another.
With this in mind, this paper provides an initial exploration of Threads, contrasting it with its parent, Instagram. 
We compare user behaviour within and across the two social media platforms, focusing on posting frequency, content preferences, and engagement patterns. Utilising a temporal analysis framework, we identify consistent daily posting trends on the parent platform and uncover contrasting behaviours when comparing intra-platform and cross-platform activities. 
Our findings reveal that Threads engages more with political and AI-related topics, compared to Instagram which focuses more on lifestyle and fashion topics.
Our analysis also shows that user activities align more closely on weekends across both platforms. Engagement analysis suggests that users prefer to post about topics that garner more likes and that topic consistency is maintained when users transition from Instagram to Threads. Our research provides insights into user behaviour and offers a basis for future studies on Threads.

\end{abstract}

\section{Introduction}

Meta's new microblogging service, Threads, was launched on July 5th, 2023. It became the fastest-growing 100 million consumer product within five days of its launch.
In contrast to prior new social networks, Threads requires an Instagram account, thus enabling rapid social bootstrapping for Threads. Threads' design allows users to follow selected or all of their Instagram follow-network already on Threads. This relationship between Threads and Instagram produces a distinct form of attention migration attributed to the unique ``parent-child''. 



This mass sign-up on Threads offers a unique perspective to characterise the discourse of early adopters on Threads. More importantly, it is possible to track these early adopters on the parent platform and offer a comparative analysis of users' behaviors to the new platform.
This is motivated by prior work that has revealed that users tend to follow varying communication norms on different platforms.
For example, users share more negative sentiment and work-related content on Twitter than on Instagram~\cite{manikonda2016tweeting}. Similarly, Instagram users tend to utilise profile images with lower smile scores compared to Twitter users~\cite{zhong2017wearing}, and they share more content on weekends than on weekdays~\cite{mytweetlim2015}.
In contrast, in the case of user migration from Twitter to Mastodon, it did not change any topical preference of users~\cite{he2023flocking}. 
We argue that the identification of such communication patterns is important for the subsequent research on user modelling and discourse analysis~\cite{kwak2010twitter}. 

Given the lack of research on Threads, here we present the first analysis of early adopters of Threads, and compare their discourse with those users on Instagram at the same time. This analysis will provide a first benchmark of Threads communication norms for future research. Specifically, we answer the following research questions:

\pb{RQ1:} What are the differences in posting behavior on Instagram vs.\ Threads? (Section~\ref{sec:RQ1}).

\pb{RQ2:} What are the content differences between Instagram and Threads, at an overall platform level? (Section~\ref{sec:RQ2}).

\pb{RQ3:} For an individual user, what factors affect the change of topic in posts? Do these factors also impact users' selection of topics when they migrate from Instagram to Threads? (Section~\ref{sec:RQ3}).




Our analysis shows that Threads and Instagram have key differences in the number of active users on weekdays during the early days of Threads' launch. In addition, there we find a difference in topical focus, with topics like \texttt{threads}, \texttt{twitter}, and \texttt{trump} being more prevalent on Threads as compared to Instagram. Finally, we show that users who consecutively write on similar topics have lower feedback on their posts than those who write on different topics in their consecutive posts.


\section{Related Work}

We first discuss the prior work on multi-social platform comparisons and the platform migration of users.


\pb{Multi-social network analysis.} In order to conduct a comparative analysis of the same users' activities across many platforms, prior studies employ algorithms to establish connections between users' accounts on other platforms~\cite{li2018matching}. 
Additionally, they utilise biography websites~\cite{zhong2017wearing} and examine self-mentions within the material across several platforms~\cite{han2016cross}.
These methods might bring different selection biases among users who utilize biography websites or self-report their personal information.
In contrast, the direct connection between Threads and Instagram obviates the need for intermediary services, simplifying the process of cross-platform analysis. We posit this offers a powerful ground truth dataset.

There have been prior studies that focus on analyzing cross-platform behaviours. 
Studies have found that users exhibit differences in self-description~\cite{mytweetlim2015} and image usage~\cite{zhong2017wearing} in profiles, alongside topic preferences~\cite{manikonda2016tweeting} and activity time~\cite{mytweetlim2015}.
Instagram, for instance, tends to be used more during leisure than work hours~\cite{mytweetlim2015}. 
The engagement for similar content also varies across different platforms~\cite{manikonda2016tweeting,farahbakhsh2016characterization}. 
User activities also differ in multi-social media platforms in terms of usage time~\cite{kumar2011understanding}.
Moreover, there is a noticeable inclination among users to prioritize certain platforms over others when disseminating similar content~\cite{farahbakhsh2016characterization,mytweetlim2015}.
Based on these previous findings, we aim to determine whether Threads exhibits comparable activity levels and topical preferences to its parent, Instagram. To the best of our knowledge, this is the first work to measure user activity on Threads.

\pb{Platform Migration.} Several prior works have studied the migration of users to different social media platforms. There are two types of migration: permanent migration and attention migration~\cite{fiesler2020moving,jeong2023exploring}.
\cite{newell2016user} explore the push and pull factors that trigger migration on Reddit.
There has also been prior work studying attention migrations from Twitter to Mastodon~\cite{jeong2023exploring}, and comparing the activities of the same users after migration~\cite{he2023flocking}. 
A key contribution of our work is the identification of a third type of migration, which we term \emph{parent-to-child migration}. This is where an existing platform spawns a new service, as is the case with Threads, which was borne out of Instagram (\eg only Instagram account owners can access Threads, and Threads posts appear on Instagram timelines as a promotional tool).

To the best of our knowledge, the closest study to our own is~\cite{jeong2023user}. This work explored the characteristics differentiating Twitter users who transition to Threads from those who stay behind. 
However, this study does not investigate the specific activities of the users on Threads, and its scope is limited to a few thousand individuals.
In our study, we utilize a dataset comprising millions of users on Threads.
Furthermore, all of these previous investigations~\cite{jeong2023exploring,jeong2023user,he2023flocking} focus on functionally similar platforms.
In our research, Threads and Instagram represent distinct platforms that face a special migration with a novel online platform~\cite {fiesler2020moving}.
To the best of our knowledge, this represents the first study to amalgamate data from both Instagram and Threads for a comparative analysis of user engagement.


\section{Dataset}\label{sec:data}

We now describe our dataset collection, pre-processing, and statistics in the following sections.


\subsubsection{Threads}

We use the threads-net API\footnote{https://github.com/dmytrostriletskyi/threads-net} to gather Threads data. We use the fact that each Threads account is associated with an auto-generated numeric ID. 
Starting from August 14th to September 4th, we iterate over all integers from 0 to 12 million to identify any account with IDs in this range. This approach helps us overcome the limitation of searching the Threads platform for specific keywords and provides an extensive search range for early Threads adopters. For each Threads account, we collect the 25 most recent threads posts with their details from that account (25 is the limit imposed by the API). 
From September 4th to September 13th, we further extend this data by snowballing the identification of additional users, and collecting their posts, based on the reposted threads (\ie, users sharing other users' content, similar to retweets on Twitter).
In total, there are 1,253,438 Threads accounts with 4,716,626 posts. 
There are 582,459 (46.47\%) users without any threads, 543,522 (43.36\%) users with fewer than 25 threads, and 127,457 (10.16\%) users with 25 threads. We also find 212 threads created before the Threads launched (which we exclude from the dataset). 
A possible explanation could be the presence of test accounts on Threads that Threads developers used to test the application's functions before its launch.



\subsubsection{Instagram}
Threads and Instagram share the same username, thus we search for all the usernames from the Threads dataset on Instagram. We use the CrowdTangle API\footnote{\url{https://crowdtangle.com}} to retrieve all corresponding Instagram accounts and their posts from May 5 to September 13, 2023. 
In total, we collect 683,168 accounts with 10,862,421 posts on Instagram based on the user list on Threads. Note, we fail to collect any Instagram accounts that have their profiles set as private, or the accounts that have changed their usernames.



For better comparative analysis with Threads, we split the Instagram data into two parts: \one Before Threads launch from May 4th to July 4th with 5,215,434 posts from 575,204 users;
and \two After Threads launch from July 5th to Sept 9th with 5,646,987 posts from 587,941 users. This segmentation helps compare users on two platforms after they start posting on Threads. There are 479,877 common users in the two datasets.




\pb{Data pre-processing.} 
We undertake several pre-processing steps.
On Instagram, the number of posts by each user is not normally distributed. The distribution of posts per user is ($\mu= 9.6, median = 4, min = 1, max = 4,942$), ($\mu= 9, median = 4, min = 1, max = 4,016$) for before and after launch data segment, respectively. Thus, we exclude any users with posts more than 0.99 percentile of number of posts per user distribution. 
This covers 83 accounts for Instagram before Threads launched and 89 after Threads launched. We also exclude the same users on Threads to ensure that the comparison is conducted on the same set of users. For RQ 1, this gives us 7,856,931 posts from 819,043 users on Threads, 4,207,344 posts from 567,179 users on Instagram before Threads launch, and 4,584,317 number of posts from 582,074 users on Instagram after Thread launch.

For the topical analysis, we exclude reposts to avoid repeat counting topics on Threads. This covers 20.83\% of Threads' posts. Note, Instagram does not have a repost feature. Hence, we do not apply this criteria on Instagram. 
On Threads, 2.88\% posts do not contain any text (\ie they only contain images), preventing us from performing topical analysis. Thus, we also exclude these.


Moreover, in the Instagram dataset, 9.38\% of images have embedded text in the images, and these parts of the text have a median length of 18 words. For these posts, we combine the text in the images (CrowdTanlge provivdes this text) with the users' added caption. We do so because Instagram is an image-centric platform, and if users have to share a text-based message such as a quote, they sometimes share it the image~\cite{manikonda2016tweeting}. 

Finally, we perform the following text pre-processing on the remaining Threads and Instagram posts' text:
\one We only include English language osts, filtered by langdetect library.\footnote{\url{https://pypi.org/project/langdetect/}} 41.47\% and 41.82\% of total posts are in non-English languages on Threads and Instagram, respectively. 
\two We remove all hyperlinks. 
\three We remove any mentions for users (@username).
\four We remove all non-alphanumeric characters, including emojis, punctuation, and special symbols (\eg \#, @, \$).
This leaves 3,533,095 posts from 556,730 users on Threads, 2,471,012 posts from 421,426 users on Instagram before Threads launch, and 2,673,157 number of posts from 436,223 users on Instagram after Thread launch. 

\section{RQ1: Time-level analysis}\label{sec:RQ1}


Using the above dataset, we first characterise the posting activities on the two platforms. We present an analysis of activities divided into weekdays and weekends. 


\subsubsection{Overview of Timeline}

\begin{figure}[t!]
    \centering
    \includegraphics[width = 0.48\textwidth]{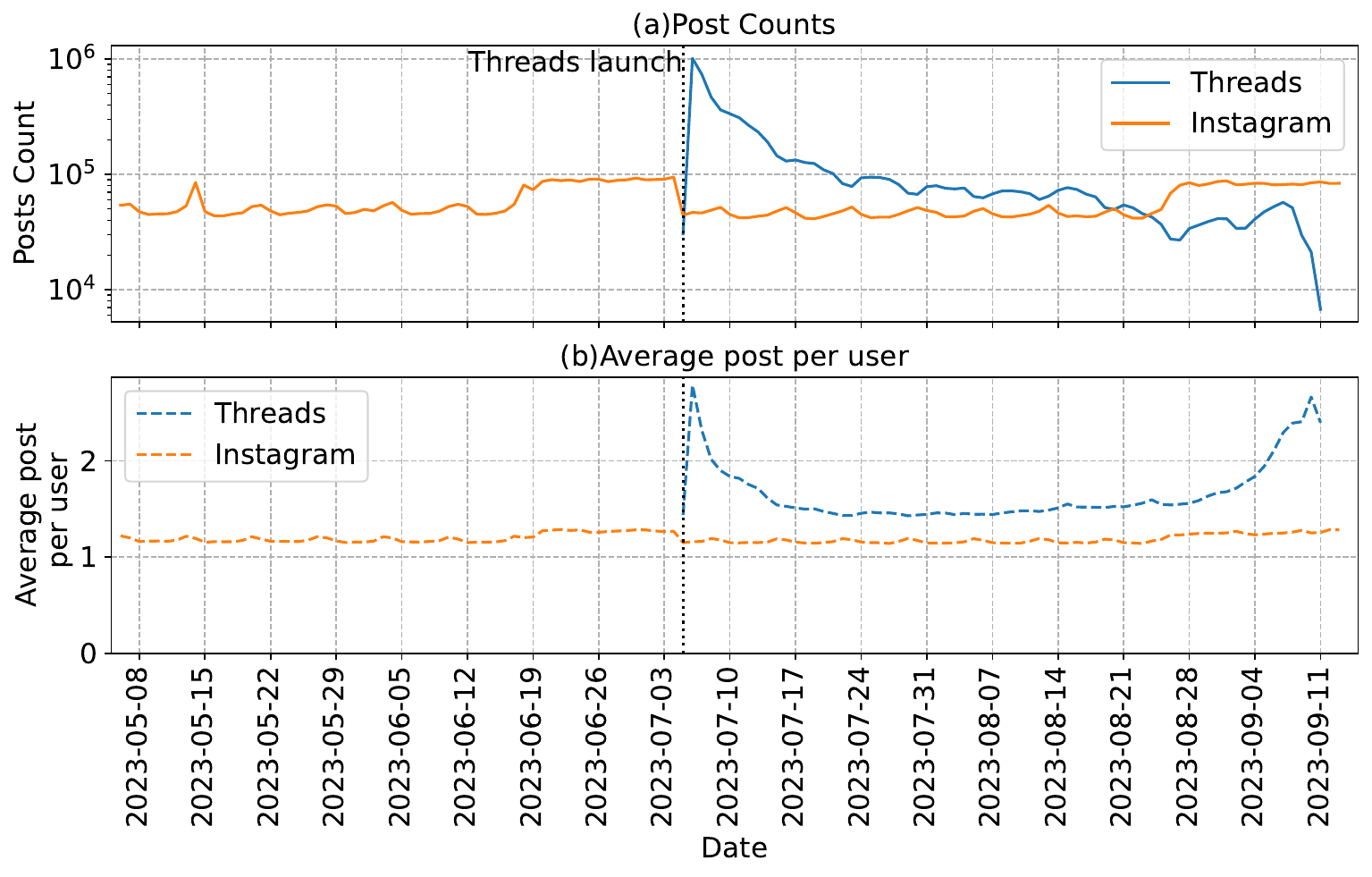}
    \caption{(a) Daily volume of posts on Instagram and Threads (b) Daily average of posts per users}
    \label{fig:ig_threads_timeline}
\end{figure}


\begin{figure}[t!]
    \centering
    \begin{subfigure}[b]{0.48\columnwidth}
        \includegraphics[width =\textwidth]{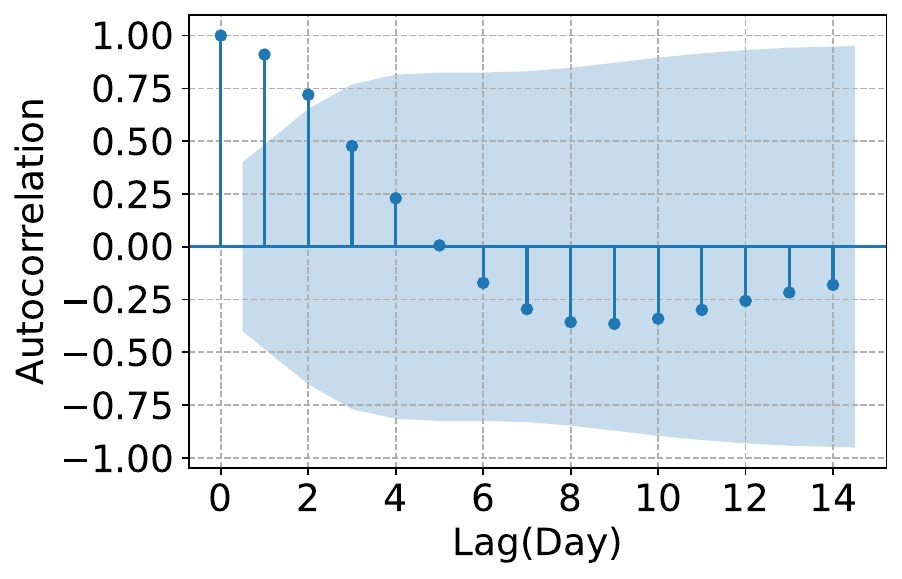}
        \caption{Instagram}
        \label{fig:ig_autocorrelation}
    \end{subfigure}
    \hfill
    \begin{subfigure}[b]{0.48\columnwidth}
        \includegraphics[width = \textwidth]{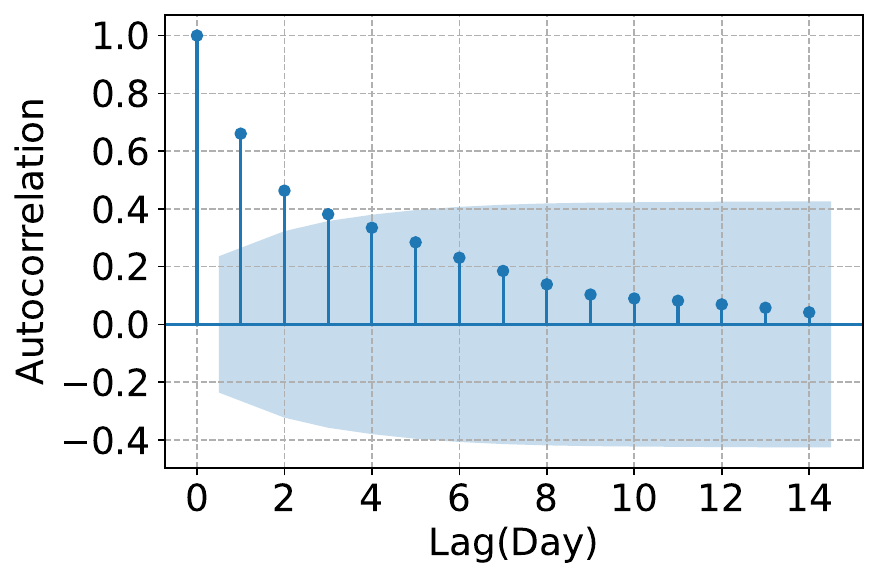}
        \caption{Threads}
        \label{fig:threads_autocorrelation}
    \end{subfigure}
    \caption{Autocorrelation for daily post counts: (a) Instagram (May 4th to Sept 9th.) (b) Threads (July 5th to Sept 9th.)} 
    \label{fig:autocorrelation}
    \vspace{-0.5cm}
\end{figure}

We start by measuring the number of posts per day per user.
Figure~\ref{fig:ig_threads_timeline} shows the (a) total post counts (solid line) and (b) the average post per user (dashed line) on two platforms daily from May 5 to September 13.
The orange line shows the data for Instagram, while the blue line shows the data from Threads, which is from its launch date.
The grid starts from the first Monday, May 8, in the dataset, and the interval is 7 days.

On Threads, the daily average number of posts per user is 1.65 compared to 1.19 on Instagram.
We see a significant peak in activity on Threads on the second day of its launch though. To gain deeper insights into the temporal patterns,, we perform auto-correlation analysis.
Specifically, we check if per-day post counts, on each platform, has a correlation with the posts counts on the following day. 


Using the daily post count as a time series, we calculate the autocorrelation function (ACF) separately for each platform. Figure~\ref{fig:autocorrelation} shows the autocorrelation function plots of Instagram and Threads.
The x-axis represents the lag in the number of days, and the y-axis represents the correlation coefficient. The blue area represents the confidence bands with 95\% confidence intervals.
Correlation coefficients outside these bands are statistically significant. The ACF plots reveal autocorrelation in the post count time series data for both platforms. Figure~\ref{fig:autocorrelation}(a) shows correlation is significant for one (y=0.90) and two (y=0.81) days. Albeit, the correlation decreases on the 2nd day. 
Similarly for Threads, Figure~\ref{fig:autocorrelation}(b) shows the correlation is statistical significance of one (y= 0.66) and two days (y=0.46).

These findings suggest that activities on both platforms \emph{do} have auto-correlation, and the number of posts on preceding days can predict the number of posts on the following day. Moreover, the correlation is higher on Instagram as compared to Threads, potentially showing that Threads nascent activities are still unpredictable.

\subsubsection{Diurnal Post Activity}


We next examine the per-hour differences between the two platforms. For this, we separate the weekends and weekdays. In addition to the posts counts, we look at the accumulated number of users who have posted during each hour. Past research on other platforms --- Twitter, Facebook, and Google+ --- indicates that users exhibit varying activity patterns on different social media platforms~\cite{mytweetlim2015}. We aim to test if post volumes match on both platforms at a given time during the day. 


\begin{figure}[t!]
    \centering
    \begin{subfigure}[b]{\columnwidth}
        \includegraphics[width =\textwidth]{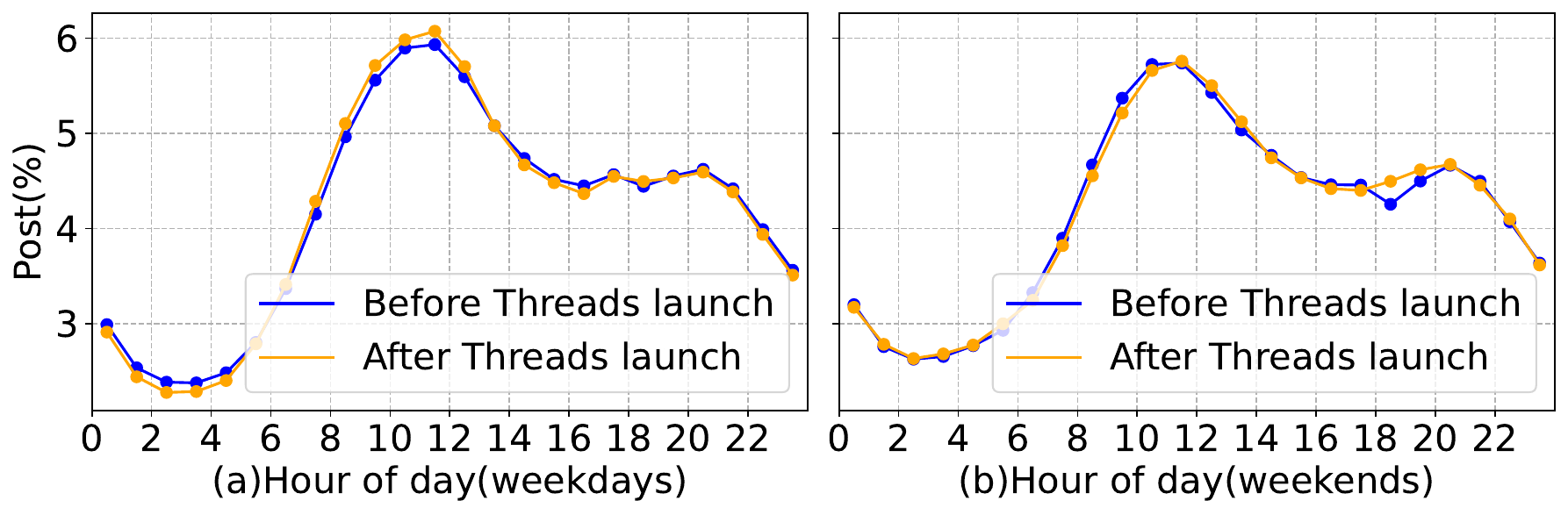}
    \end{subfigure}
    \hfill
    \begin{subfigure}[b]{\columnwidth}
    \includegraphics[width = \textwidth]{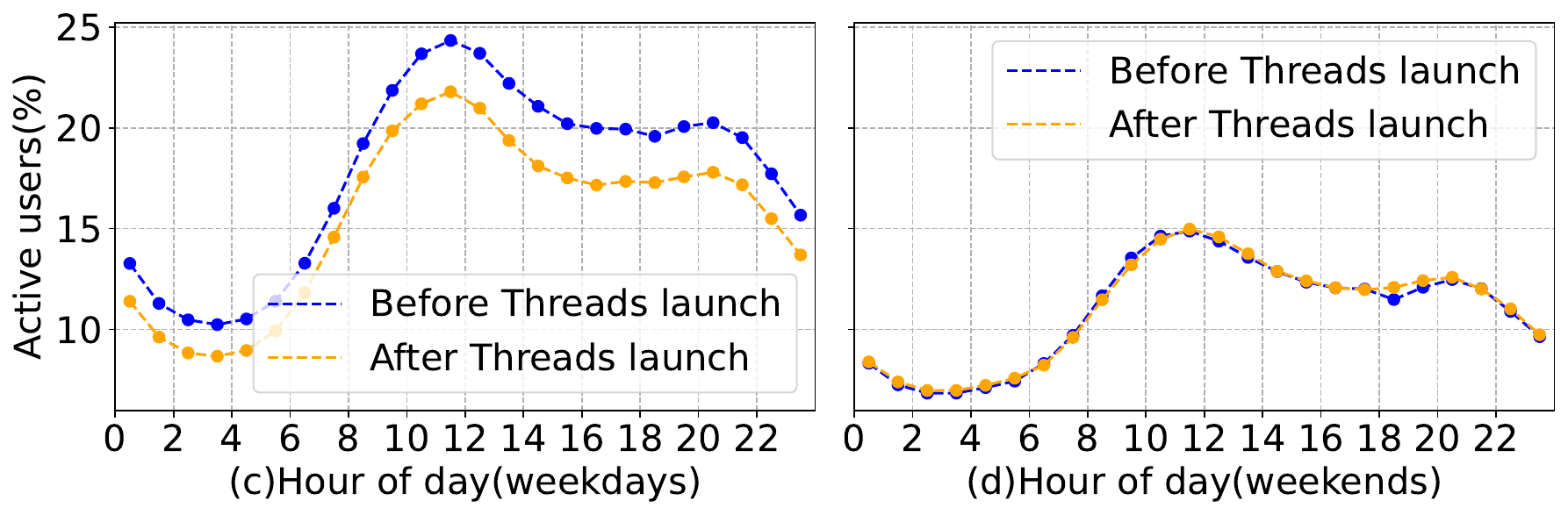}
    \end{subfigure} 
    \caption{Compared Instagram before and after Threads launch: (a) and (b) show the percentage of accumulated posts in each hour on weekdays and weekends, respectively. (c) and (d) show the percentage of accumulated active users in each hour on weekdays and weekends, respectively. Times are converted to EDT zone.}
    \label{fig:ig_compare}
    \vspace{-0.5cm}
\end{figure}






\pb{Instagram Before and After the Threads Launch.} 
We partitioned the Instagram datasets (before and after Threads launch) into weekdays and weekends, separately.
Then, we split each day into 24 hourly slots and count the number of posts and users in each hour, across the Instagram datasets (before and after Threads launch). 
Figure~\ref{fig:ig_compare} shows the number of posts ((a) and (b)) and users ((c) and (d)) as a percentage of the total posts and users in any given hour.
Note, using the percentage gives a normalised representation of the data. The blue lines represent the Instagram data for before Threads launch period and the orange lines represent the Instagram data after Threads launch period.

First, we look at the posts in two periods for weekdays (Figure~\ref{fig:ig_compare}(a)) and weekends (Figure~\ref{fig:ig_compare}(b)). We observe that the accumulated percentage of posts on the two platforms have similar patterns on weekends and weekdays. We note that the posts volume on the platform starts to increase from early morning (05:00) and reaches its peak at 11:00. It then begins a downward trend with marginally stable activities from 16:00 until 20:00. Similar patterns hold for both weekends and weekdays. 

However, in terms of users, we see contrasting results, especially on weekdays. Figure~\ref{fig:ig_compare}(c) shows that the overall trends of active users in hourly windows are similar on Instagram (both after and before the Threads launch). The number of active users peaks between 11:00 -- 12:00 AM. Interestingly, the proportion of active users has dropped after the Threads launch with a difference of 2.5\% at the peak hour.
On weekdays, overall proportion of active Instagram users is 17.7\% before the Threads launch, and 15.6\% on after. 
In contrast, the proportion of active users does not change on weekends (average of 11.0\%) before and after, as shown in Figure~\ref{fig:ig_compare}(d). 



For Instagram, we find that during the weekends there is no difference in terms of active users and the posts count in an hourly proportion before and after the Threads launch. 
However, there are differences in terms of user counts. 
We find that the proportion of active users during weekdays has reduced after the Threads launch. However, there is no change in terms of the post pattern on the platform. Overall, we observe a reduction in the proportion of users during the weekdays, but there is no change in the proportion of post counts in hours. 

\begin{figure}[t!]
    \centering
    \begin{subfigure}[b]{\columnwidth}
        \includegraphics[width =\textwidth]{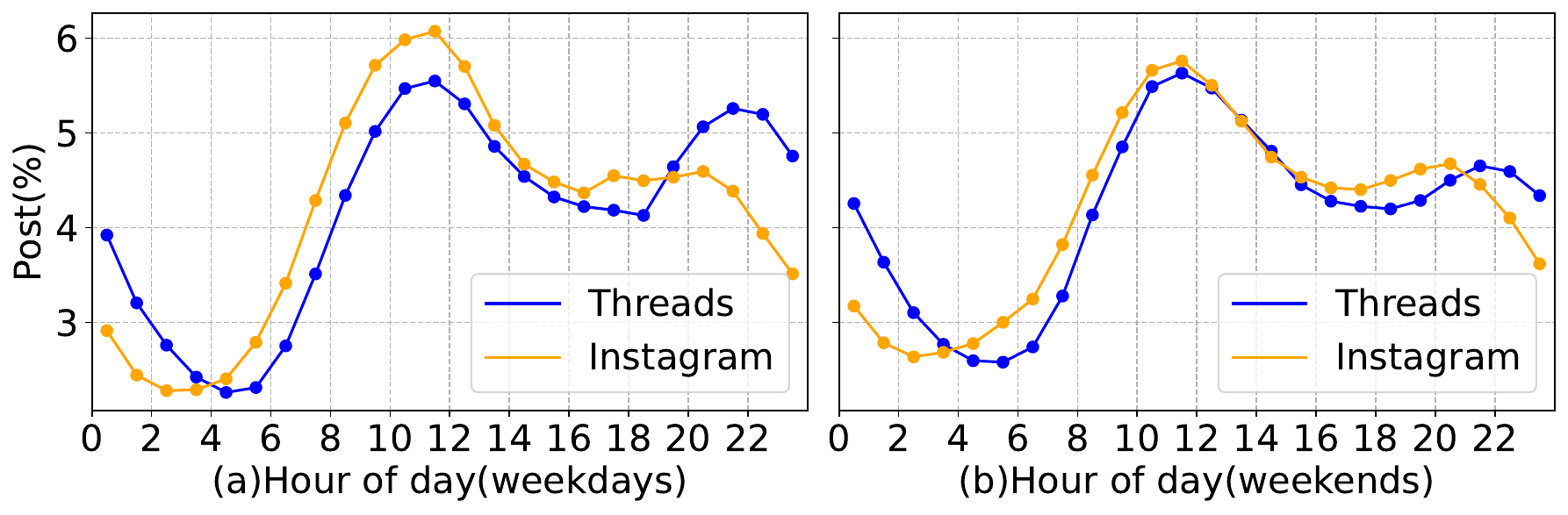}
    \end{subfigure}
    \hfill
    \begin{subfigure}[b]{\columnwidth}
    \includegraphics[width = \textwidth]{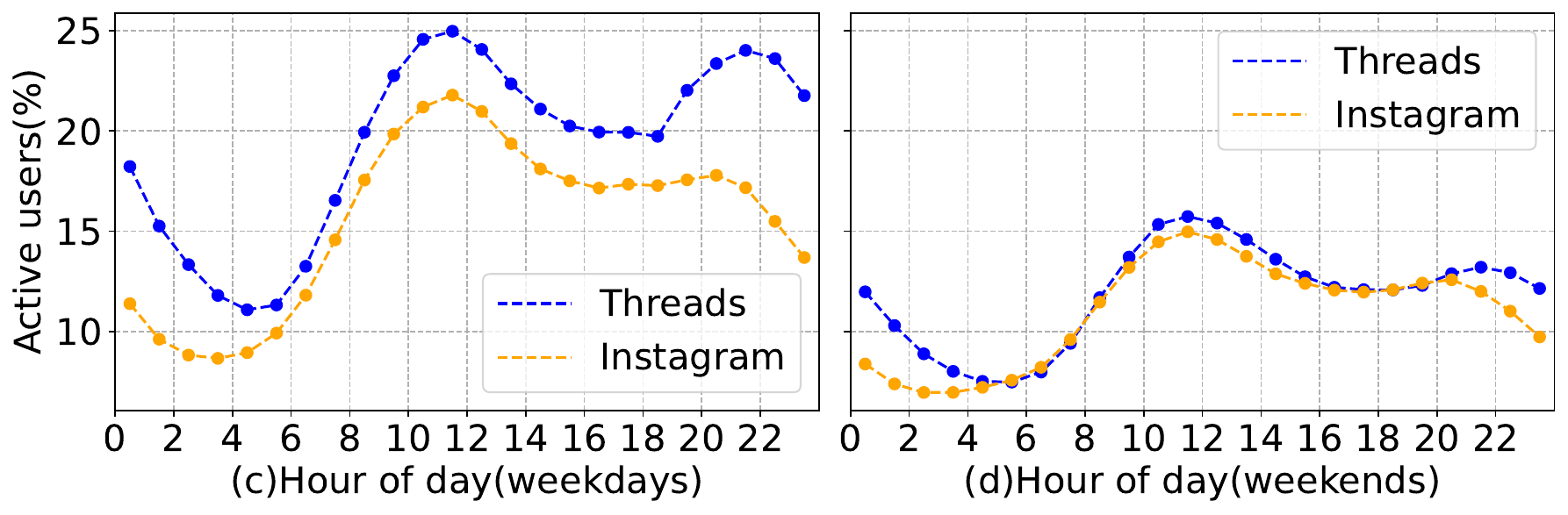}
    \end{subfigure} 
    \caption{Compared Threads and Instagram after Threads launch: (a) and (b) show the percentage of accumulated posts in each hour on weekdays and weekends, respectively. (c) and (d) show the accumulated active users in each hour on weekdays and weekends, respectively. Times are converted to EDT time zone.}
    \label{fig:threads_hour}
    \vspace{-0.1cm}
\end{figure}

\pb{Comparison between Instagram and Threads.}
Next, we compare the hourly activity patterns on Threads with Instagram. For this, we only consider the Instagram data after the Threads launch. 
Figure~\ref{fig:threads_hour} shows the hourly percentage of posts and active users on Threads and Instagram (after Threads launch). It is further split in weekdays (Figure~\ref{fig:threads_hour}(a)) and weekends (Figure~\ref{fig:threads_hour}(b)). 
We see that the proportion of posts on both platforms follows a similar trend during most of the accumulated 24 hour window. The activities drop on both platforms from midnight to early morning (up to 5:00 AM) and then reach the maximum at 11:00 (Instagram: 22.0\%, Threads: 25.0\%). Interestingly, the activity levels start to differ from 18:00 hours though.
After this time, the overall percentage of posts drops on Instagram, however, on Threads, it achieves another peak at around 21:00 with a positive change of 4.3\% from 18:00.
On weekends, both platforms follow a largely similar pattern of active users and reach a peak at 11:00 hours with 15.7\% and 15.0\% for both Instagram and Threads, respectively. 

Moreover, for the hours with similar trends on both platforms, we observe a lag between the two platforms, with Instagram leading Threads as shown in (Figure~\ref{fig:threads_hour}(a)) and weekends (Figure~\ref{fig:threads_hour}(b)).
To understand this time lag, we perform a cross-correlation function (CCF) on the percentage of activities. We take each data series (Instagram and Threads) and use the CCF to identify lags of one series that might be useful predictors of another. We take the activities on Instagram as $x_t$ and Threads as $y_t$, separately on weekdays and weekends.
The results show that the most dominant cross-correlations occur at a lag of -1 hour
both on weekdays (cross\_correlation=0.919) and weekends (cross\_correlation=0.961).
The results of the cross-correlation analysis fall outside the bounds of the 95\% confidence interval (from -0.4 to 0.4). This indicates statistical significance. The results mean that on the aggregate post activities time series the activities on Instagram lead the activities on Threads by 1 hour. 


We then compare the count of active users in each hour.
Figure~\ref{fig:threads_hour}(c) and Figure~\ref{fig:threads_hour}(d) plot the active user counts accumulated per hour on weekdays and weekends separately on Instagram and Threads.
The blue dashed-line is higher than orange dashed-line in Figure~\ref{fig:threads_hour}(c).
This means there are more users active on Threads at any given hour during weekdays as compared to Instagram.
Users are counted for each hour they post. 
We further measure the average number of different hours where users posts on each platform. On weekdays, the average number of different hours a user posts is
4.65 and 3.74 for Threads and Instagram, respectively. Given that we have more users on Threads, as compared to Instagram, after the launch and a higher average number of hours per user, the higher number of active users is justified. 
Compared to Instagram, a greater percentage of users are active on Threads during the night from 21:00 to 4:00 on both weekdays (38.47\%) and weekends (11.73\%). The biggest difference in hours is 22:00 on weekdays and 0:00 on weekends.
Threads has a higher percentage of active users at night than Instagram. 

Overall, we find that within Instagram, the proportion of post distribution in the hours window does not differ before and after the Threads launch. However, the proportion of active users in hours changes before and after the Threads launch. When being compared with Threads, the Instagram volume of posts on Instagram leads Threads by one hour. However, the overall pattern of content generation and users being active remains similar on both platforms, with the maximum number of users being active around noon, and hence the most posts are also shared at the same time.

\section{RQ2: Content-level analysis}\label{sec:RQ2}


We now analyse the topics 
on both platforms. We also look at the temporal evolution of topics to understand whether users continue a similar discourse on both platforms.

\begin{figure}[t!]
    \centering
    \includegraphics[width = 0.5\textwidth]{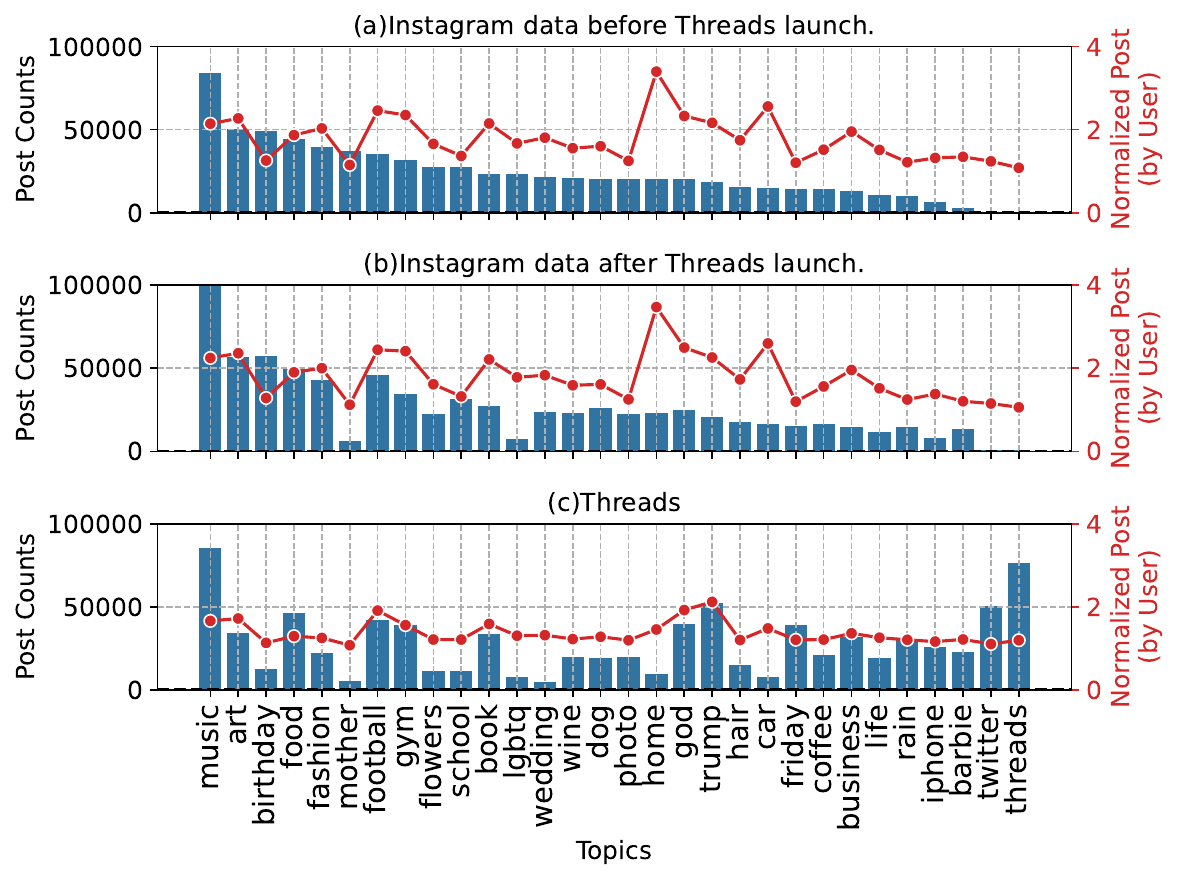}
    \caption{Posts count for the top-30 topics (left-y) and normalised posts per user(right-y). (a) Instagram before Threads launch. (b) Instagram after Threads launch. (c) Threads. The x-axis is ranked based on the number of posts on Instagram before Threads launch.}
    \label{fig:topics_barchart}
    \vspace{-0.5 cm}
\end{figure}


\subsubsection{Topical Discussions} \label{sec:topic_discussion}
We first look at the most common topics on each platform, and their weekly trends.
We use BERTopic~\cite{grootendorst2022bertopic}, which employs sentence transformers and c-TF-IDF to generate topics. We combine the posts from both platforms to create one corpus to ensure consistency and coherence in topic comparisons~\cite{he2023flocking}.
Using a threshold of at least 3,500 posts per topic, we 
identify 206 topics across the combined corpus. These topics cover 50.27\% of posts in the data. 

\pb{Topic Volume.} We first look at the most common topics within each platform. 
We see that the prominence of topics varies across the two platforms.
Figure~\ref{fig:topics_barchart} presents the absolute number of posts (bars) and the normalized number of posts per user (red line) for the top 30 topics. The x-axis shows the topics ranked in order of their post count in the Instagram dataset before the Threads launch. 
Normalization is performed by calculating the number of users engaging in a given topic. 
Figure~\ref{fig:topics_barchart}(a) and (b) have similar trends in terms of both the total number of topics and the normalized result. 
Some topics vary on Instagram before and after Threads launch in the total number of posts. 
These topics are related to specific events occurring during that period, such as \texttt{mother, barbie (film)} and \texttt{lgbtq}, which refer to Mother's Day, film-release and LGBT pride, respectively. 

Figure~\ref{fig:topics_barchart}(c) presents the topics on Threads.
When compared with Figure~\ref{fig:topics_barchart}(b), we notice that topics about art (\texttt{music} and \texttt{art}), food (\texttt{food, coffee} and \texttt{wine}), and sports (\texttt{football} and \texttt{gym}) are popular on both Instagram and Threads.
However, Threads also has distinct topics from Instagram.  
Topics on social media applications (\texttt{Threads} and \texttt{Twitter}) are more discussed on Threads. 
Politics(\texttt{trump}) is another topic discussed more on Threads than on Instagram.
However, discussion patterns on the two platforms display a different trend.
A total of 24,659 (4.43\%) users wrote about topic \texttt{trump} on Threads. Yet there are only 9,113 (2.09\%) users writing about \texttt{trump} on Instagram. 
Interestingly, each user on Instagram is more active though, with 2.26 posts per user about \texttt{trump} vs.\ 1.43 posts per user on Threads.
Furthermore, Threads has fewer posts on daily lifestyle topics like fashion (\texttt{fashion} and \texttt{hair}) and pets (\texttt{dog}) as compared to Instagram, where these topics are popular~\cite{hu2014we}. Moreover, users prefer to share festivals and anniversary topics such as \texttt{birthday} and \texttt{wedding} on Instagram. 
Instagram has a higher total number of posts on these topics than Threads. 
3,850 users post \texttt{wedding} relevant content on Threads, and 12,855 users post wedding on Instagram. 925 (7.20\% of 12,855) users post this topic on two platforms. 11,111 users post \texttt{birthday} on Threads, and 44,677 users on Instagram. There are 1,914 (4.28\% of 44,677) common users who post this topic on two platforms.
Finally, we also notice a topic \texttt{AI} outside Figure~\ref{fig:topics_barchart}.
10,509 (1.89\%) users engaging in discussion on \texttt{ai} with an average of 2.12 posts per user on Threads.
Only 3,282 (0.75\%) users discussing \texttt{ai} on Instagram.
It indicates a potential interest in AI-relevant topics on Threads that might not be as prominent on Instagram. 

Overall, we note that while there are topical similarities on the two platforms, users on Threads post less on fashion-related topics and more on topics related to politics and technology than Instagram.

\begin{figure*}[t!]
    \centering
    \includegraphics[width = 0.9\textwidth]{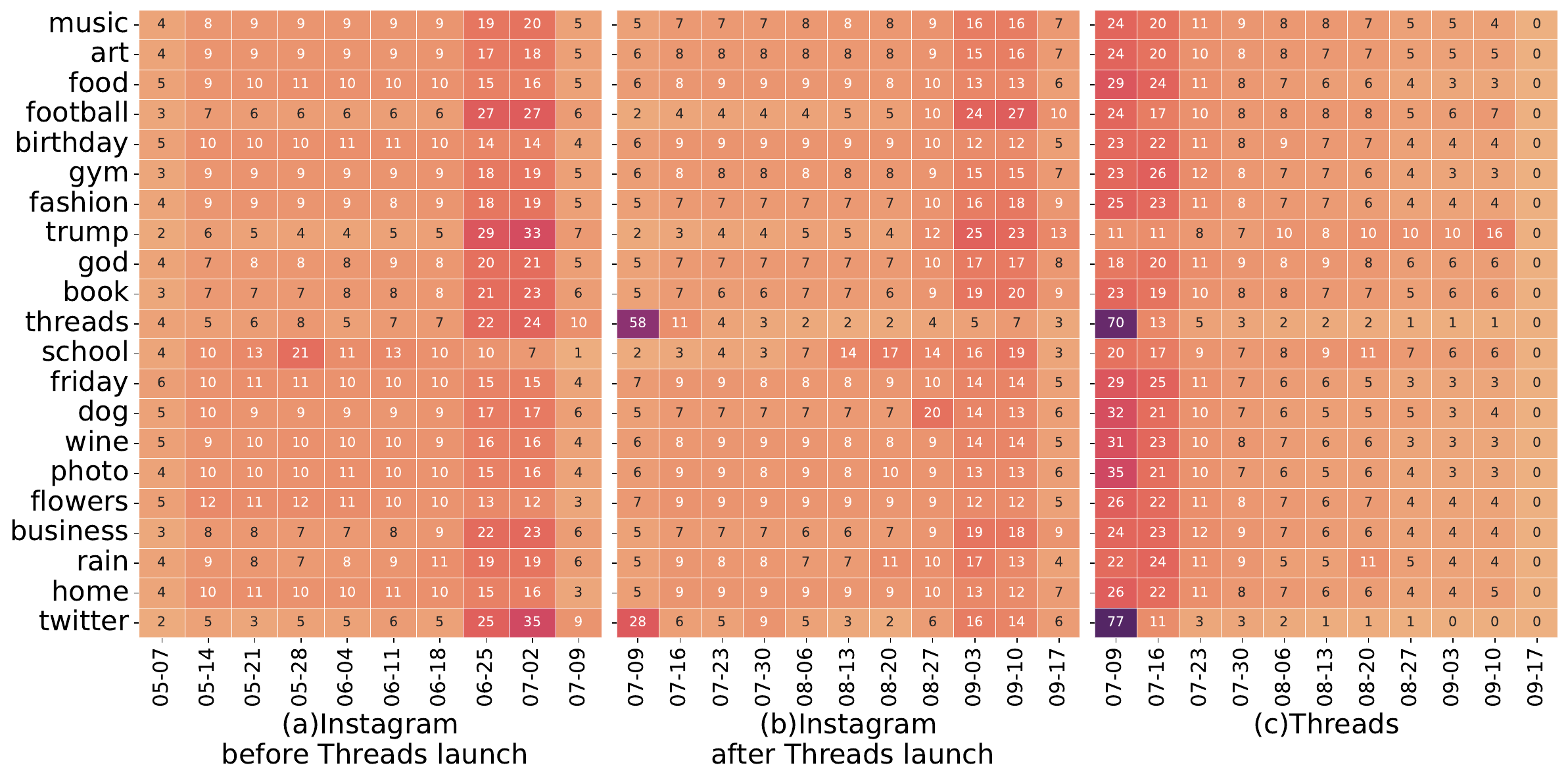}
    \caption{The 3 heatmaps for top 20 topics on (a)Instagram before Threads launch, (b)Instagram after Threads launch and (c)Threads. Normalise by the number of each topic/total number on this platform.}
    \label{fig:topic_by_time}
    \vspace{-0.3cm}
\end{figure*}


\pb{Topics Over Time.} 
We then check the temporal variations of the top 20 topics on two platforms. To do so, we compare the weekly volume of each topic over the duration of dataset. 
Figure~\ref{fig:topic_by_time} shows the normalised number of posts for a given topic over time. To be more precise, for each topic we calculate the total number of posts across all the dataset, and divide the weekly count by the total number. Hence, giving us a normalised change over time, and each row will sum up to 100 percent. 
The x-axis ticks show all the Sundays.
Sunday here is the end of the week.
Figure~\ref{fig:topic_by_time}(a), (b) and (c) 
share the same colour scale.
On Threads, we notice that the discussion about \texttt{threads} and \texttt{twitter} on Threads mainly stem from the first week after Threads launch.
It also spurs discussion on these two topics on Instagram in the same week. However, combining the results from the previous section, the volume of this discussion is relatively small compared to Threads.
On both platforms, the discourse on \texttt{threads} and \texttt{twitter} rapidly dissipates. 
Most topics on Threads have a decreasing trend, except \texttt{trump}.
The number of posts on \texttt{trump} keeps a stable trend, slightly increasing in two months.
It hints at a potential interest in political discussion on Threads.

\section{RQ3: Topic-consistency exploration}\label{sec:RQ3}
Finally, we analyze the impact of the feedback received by posts in individual users' topical choices. We explore whether users' topical choices change as they move from Instagram to Threads. As such, we first define the term topic consistency and then explore the changes in users' topic consistency within and across the platform.


\subsection{Topic-consistency and feedback} \label{sec:definition of topic-consistency}

We define topic consistency as continuing the same topic in two consecutive posts. We further split this into two levels to examine topic consistency at the intra- and inter- platforms scale. 


\subsubsection{Intra-platform topic-consistency}

If two consecutive posts by a user on a single platform share the same topic, the preceding post in time will be labeled as having Intra-platform topic consistency.

\subsubsection{Inter-platform topic-consistency}
Inter-platform topic consistency measures a user's continuation of the same topic from one Instagram to Threads.


To explore the topic consistency of users when migrating from Instagram to Threads, we inspect posts for each user.
We first extract the topics of $n$ recent posts on Instagram before the launch of Threads (the specific number of posts will be discussed as a threshold in Section~\ref{sec:user-level}). 
We then extract the topics of $n$ posts on Threads.
If there is an overlap between these two sets, the user will be regarded as having inter-platform topic consistency.
However, it is difficult to select the optimal threshold ($n$) for the number of posts to calculate this overlap. 
Hence, we experiment with three thresholds (\ie, the first or latest 1/2/3 posts to calculate the overlap) and compare the results of these three alternatives.

\subsubsection{Feedback}\label{sec:feedback}


We also measure feedback on posts, using the number of likes for each posts on both platforms. To calculate the \emph{feedback} of a post, we first calculate the mean value of the likes of all posts published before this post. Taking the mean value of the previous likes and comparing it with the likes received by the current post helps in measuring whether the new post received better or worse engagement. 
We use the obtained mean value to divide the likes count by the current post, and use Laplace smoothing to avoid a denominator of zero. 




\subsection{Intra-platform exploration}
\label{sec:post-level}

We now estimate this feedback measure's impact on topic consistency at post and user levels. We hypothesize that topic-consistent posts tend to have worse feedback since users with unpopular posts tend to focus more on their interests (narrow interests) compared with more popular users \cite{ferrara2014online}. 


For this, we first divide all the posts into topic-consistent posts and topic-inconsistent posts. From all the posts that have been assigned a valid topic from BERTopic model (same model from RQ2), there are 709,272 (28.25\%) topic-consistent posts and 1,801,559 (71.75\%) topic-inconsistent posts on Instagram, while 115482 (8.93\%) topic-consistent posts and 1178260 (91.07\%) topic-inconsistent posts on Threads. 
There are fewer consistent topics posts on Threads than on Instagram as a percentage. 

\begin{figure}[t!]
    \centering
    \includegraphics[width = 0.25\textwidth]{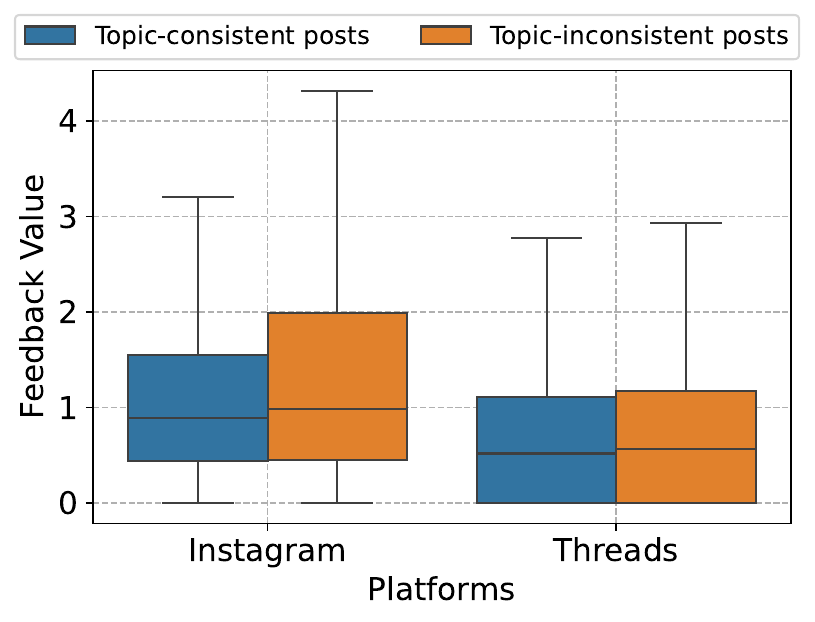}
    \caption{Box plot of feedback value for both topic-consistent posts and inconsistent posts on Instagram and Threads.}
    \label{fig:postlevel_feedback_boxplot}
\end{figure}



We now analyse the feedback differences on topic-consistent and inconsistent posts. Figure~\ref{fig:postlevel_feedback_boxplot} shows the distribution of feedback for both Threads and Instagram. On Threads, the first quartile of posts have zero feedback because some posts have no likes on the newly launched Threads.
We find that the feedback value on Instagram (median = 0.95, std = 10853.97) is higher than that on Threads (median = 0.55, std = 180.74) with the K-W test ($H (1) = 33302.96$, p-value \textless 0.001).
In general, topic-inconsistent posts (median = 0.88, std = 9637.13)  have relatively higher feedback than topic-consistent posts (median = 0.79, std = 7800.72) 
(with K-W test results: $H (1) = 770119.22$, p-value \textless 0.001 on Instagram and $H (1) = 62099.37$, p-value \textless 0.001 on Threads).

Unsurprisingly, these results indicate that topic-inconsistent posts are more likely to have better feedback, \ie users who talk about a topic that receives a high number of likes are more likely to change the topic of their next post.
This phenomenon is more common on Instagram. 
A potential explanation for this phenomenon is that users who prefer to publish posts on a wide variety of topics tend to be popular with audiences. Naturally, receiving positive feedback reinforces this message for publishers. But users who focus on same topics care little about the feedback.


\subsection{Inter-platform exploration}\label{sec:user-level}
Next, we utilize the inter-platform consistency definition (users writing similar topics on Instagram and Threads in their consecutive posts) and divide users into consistent and non-consistent groups. 
We test three different thresholds on number of consecutive posts (from 1 to 3) to define the inter-platform topic consistency. A threshold of 3 posts means that we require users to repeat a topic at least once in the 3 consecutive posts to be considered topic-consistent. A higher threshold will increase the number of topic-consistent users. Conversely, a lower threshold (1 post) will lower the number of topic-consistent users. Table~\ref{tab:thresholds_users counts} shows the number of consistent and inconsistent users with all thresholds.


\begin{table}[]
\centering
\resizebox{1.0\columnwidth}{!}{
\begin{tabular}{lccc}
\toprule
  & \textbf{Threshold =1}  & \textbf{Threshold =2} & \textbf{Threshold =3} \\ \midrule
Consistent users   &  46,051 &  84,631 & 102,807 \\ 
Inconsistent users  &  131,293 & 92,713 & 74,537  \\ \bottomrule
\end{tabular}}
\caption{The number of consistent/inconsistent users for different threshold.}
\label{tab:thresholds_users counts}
\vspace{-0.5cm}
\end{table}

\begin{figure}[t!]
    \centering
    \includegraphics[width = 0.25\textwidth]{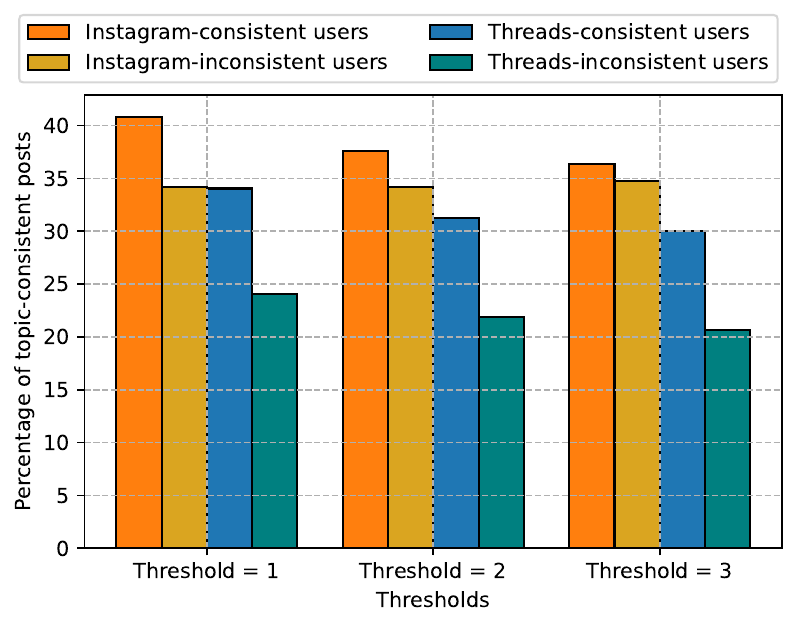}
    \caption{Percentage of topic consistency posts for consistent/inconsistent users on two platforms (with three different thresholds).}
    \label{fig:consistentpercent_bar}
\end{figure}

We first analyze the percentage of topic-consistent posts by topic-consistent and -inconsistent users. Figure~\ref{fig:consistentpercent_bar} shows these percentages for both Instagram and Threads with all thresholds. This figure shows that the percentage of topic-consistent posts (with threshold=1), on Instagram, from consistent users (40.83\%) is higher than inconsistent users (34.14\%).
The corresponding percentage on Instagram is also higher than Threads (\ie 40.83\%/34.14\% on Instagram is higher than 34.04\%/24.03\% on Threads when threshold = 1).

These results indicate that users who maintain topic consistency when they migrate from the ``parent'' platform to the ``child'' are also more likely to publish posts on the same topics consecutively. Further, the same group of users is more likely to maintain topic consistency on Instagram but change topics more frequently on Threads.

\begin{figure}[t!]
    \centering
    \begin{subfigure}[b]{0.49\columnwidth}
        \includegraphics[width =\textwidth]{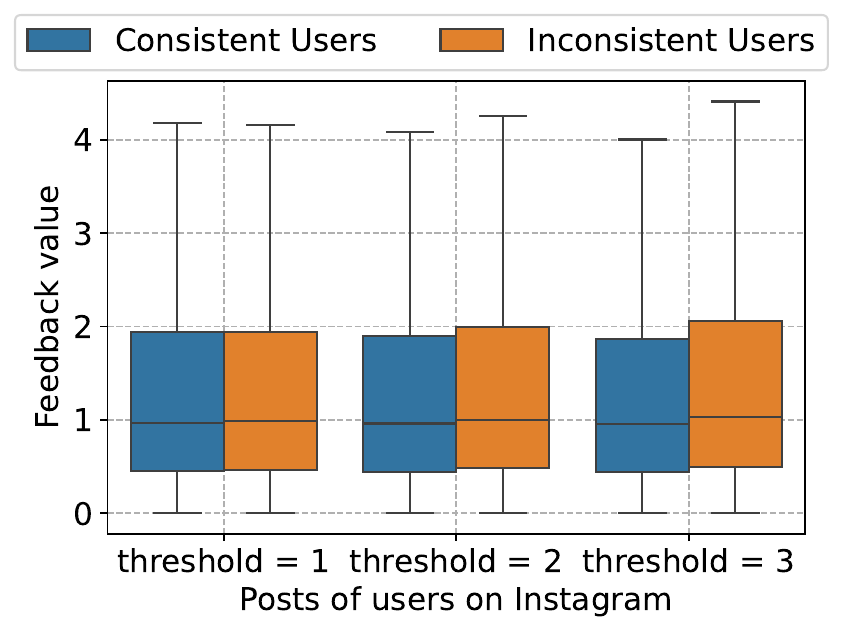}
        \caption{Feedback (Ins)}
        \label{fig:userlevel_feedback_ins}
    \end{subfigure}
    \hfill
    \begin{subfigure}[b]{0.49\columnwidth}
        \includegraphics[width = \textwidth]{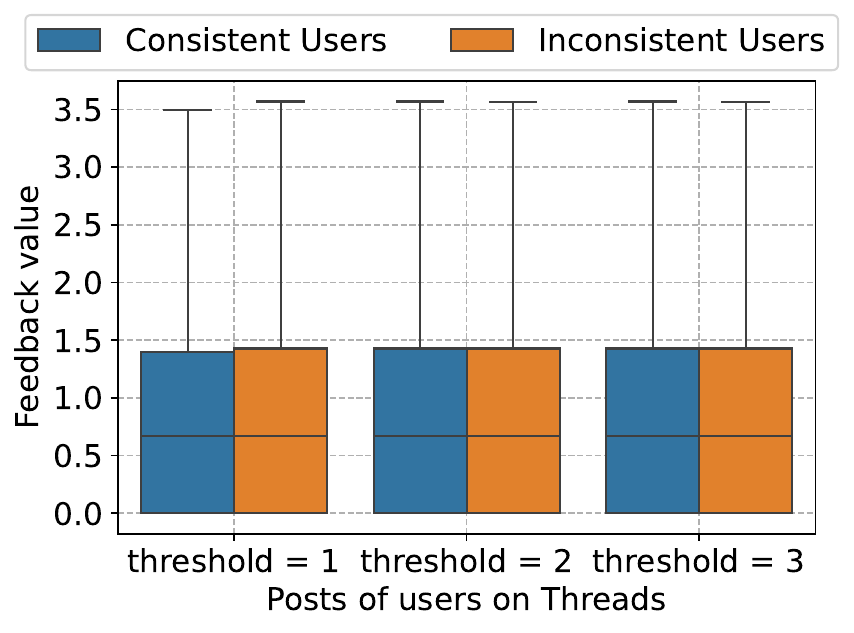}
        \caption{Feedback (Threads)}
        \label{fig:userlevel_feedback_threads}
    \end{subfigure}
    \caption{
    (a) Box plot of feedback for consistent/inconsistent users on Instagram. (b) Box plot of feedback for consistent/inconsistent users on Threads.}
    \label{fig:userlevel_boxplot}
    \vspace{-0.5 cm}
\end{figure}

We finally analyze the differences in feedback on posts from these user groups and show the distribution of feedback in Figure~\ref{fig:userlevel_boxplot}(a) and Figure~\ref{fig:userlevel_boxplot}(b) for Instagram and Threads, respectively. The median feedback count on Threads is less than 1 for both consistent and inconsistent users on all thresholds. On Instagram, the median value is close to 1 for both user types. A K-W test shows that there is no significant difference in feedback received by both user-groups, even though users' feedback on Instagram (median = 0.98, std = 15528.53) is relatively higher than that on Threads (median = 0.67, std = 118.56). These results indicate that there is little correlation between whether users maintain topic-consistency at inter-platforms scale and the feedback they receive. And generally users receive better feedback on Instagram compared with Threads.




\section{Conclusion}\label{sec:conclusion}

Our study serves as the first characterization of Threads.
We focus on the inteheritted overlaps between Threads (the ``child'') and Instagram (its ``parent'').
We have inspected the posting activities and topical discourse of the early adopters of Threads. We have also compared the same users on Instagram to record the difference in users' communication on both platforms. 

Our analysis (RQ 1) has shown that both Instagram and Threads follow a similar posting pattern across 24-hour accumulated cycles. Both platforms have peak activities at noon. However, the proportion of active users on Instagram is lower after the Threads launch, albeit the proportion of posts in the hour window remains consistent. In addition, we note that Instagram leads the activities on Threads by 1 hour. 
We have also shown that there is a difference in topical focus across the two platforms (RQ 2). Instagram has a higher percentage of topics related to lifestyle and fashion, while Threads has more topics related to politics and technology. 
Finally, we have explored the topic-consistency across platforms and compare the feedback (RQ 3). 
We find that users care more about the content itself than about the feedback their posts receive, especially on Instagram. We also notice that users who maintain topic consistency when they migrate from Instagram to Threads are more likely to publish posts consecutively on the same topics.

\pb{Limitation and Future Work.} This is the first work on Threads. However, we wish to flag several limitations mainly arising from the data collection. First, we are limited by 25 posts per user on Threads. Second, some Threads users cannot be associated with their Instagram accounts due to their privacy settings or username changes. Third, our longitudinal analysis is limited to a two month duration and cannot necessarily be generalized to the whole platform. We plan to extend our data collection and the observation period to expand our findings, and holistically characterize the users' movement from a parent to child platform. Other potential research areas include the impact on the parent's platform activities, \eg related to Instagram's e-commerce features.

\bibliographystyle{IEEEtran}

\end{document}